\documentclass[aps,prl,reprint,superscriptaddress]{revtex4-2}
\usepackage{hyperref}
\usepackage[english]{babel}
\usepackage{amsmath}
\usepackage{amssymb}
\usepackage{graphicx}
\usepackage{siunitx}
\usepackage{textcomp,xcolor}
\usepackage{textgreek}
\newcommand{\ignore}[1]{}

\begin{document}

\title{The role of band matching in the proximity effect between a superconductor and a normal metal}

\author{Loic Mougel}
\affiliation{Physikalisches Institut, Karlsruhe Institute of Technology, 76131 Karlsruhe, Germany}
\author{Patrick M. Buhl}
\affiliation{Institute of Physics, Johannes Gutenberg University Mainz, 55099 Mainz, Germany}
\author{Qili Li}
\affiliation{Physikalisches Institut, Karlsruhe Institute of Technology, 76131 Karlsruhe, Germany}
\author{Anika M\"uller}
\affiliation{Physikalisches Institut, Karlsruhe Institute of Technology, 76131 Karlsruhe, Germany}
\author{Hung-Hsiang Yang}
\affiliation{Physikalisches Institut, Karlsruhe Institute of Technology, 76131 Karlsruhe, Germany}
\author{Matthieu Verstraete}
\affiliation{Nanomat/Q-mat/CESAM and European Theoretical Spectroscopy Facility, Universit\'e de Li\`ege, B-4000 Sart Tilman, Belgium}
\author{Pascal Simon}
\affiliation{Laboratoire de Physique des Solides, UMR 8502, CNRS, Universit\'e Paris-Sud, Universit\'e Paris-Saclay, 91405 Orsay, France}
\author{Bertrand Dup\'e}
\affiliation{Fonds de la Recherche Scientifique (FNRS), B-1000 Bruxelles, Belgium}
\affiliation{Nanomat/Q-mat/CESAM, Universit\'e de Li\`ege, B-4000 Sart Tilman, Belgium}
\author{Wulf Wulfhekel}
\affiliation{Physikalisches Institut, Karlsruhe Institute of Technology, 76131 Karlsruhe, Germany}

\date{\today}

\begin{abstract}
We combine density functional theory and scanning tunneling microscopy to study the proximity effects between a bulk Ru(0001) superconductor and an atomically thin overlayer of Co. We have identified that the Co monolayer can grow in two different stackings: the hcp and a reconstructed \textepsilon-like stacking. We analyze their electronic structure from both experiments and density functional theory.  While the magnetic hcp stacking shows a weak proximity effect in combination with Shiba states and with almost no suppression of superconductivity of the substrate, the more complex \textepsilon-like stacking becomes almost fully superconducting and displays an edge state at the island rim.
We identify this edge state as a trivial state caused by a local hcp rim around the \textepsilon-core. We explain the weak proximity effect between Ru and the magnetic hcp islands by a low transparency of the interface, while the large chemical unit cell of the non-magnetic \textepsilon-like stacking lifts the momentum conservation at the interface making it transparent and causing a clear proximity effect.  

\end{abstract}

\maketitle
\noindent

The superconducting proximity or Holm-Meissner effect is a phenomenon that arises when a normal metal is placed in contact with a superconductor \cite{Holm1932}. While the normal metal can be described in the framework of Fermi liquid theory, i.e. the single electron Bloch states are filled up to the Fermi level following Fermi-Dirac statistics, in the superconductor a gap appears in the single particle spectrum near the Fermi level and electrons condense in Cooper pairs \cite{Cooper1956}. In conventional superconductors, the pairing energy is a consequence of an attractive interaction between electrons mediated by virtual phonon exchange \cite{Bardeen1957}. The Cooper pairs form a condensate, whose density reflects the superconducting order parameter. Lateral variations of the Cooper pair density inside the superconductor typically arise on the length scale of the superconducting coherence length $\xi$.

When bringing a superconductor in contact with a normal metal, Cooper pairs may be scattered into the normal conductor and unpaired electrons may be scattered into the superconductor. As in the normal metal, the attractive interaction between electrons is absent, Cooper pairs decay into single electrons after traveling a characteristic coherence length $\xi_n \neq 0$. In case the normal metal is magnetically ordered, the exchange interaction breaks the Cooper pairs, resulting in rather short coherence lengths. 
More generally, as states on both sides of the interface are described by wave functions in momentum space, they are de-localized in real space and cannot change on arbitrarily short distances.
As a consequence, when approaching the interface, the superconducting order parameter decays from its bulk value far inside the superconductor, leaks into the normal metal, and finally vanishes far inside the normal metal.
The superconducting properties are thus transferred over some distance into the normal metal, and the order parameter in the superconductor is lowered near the interface\cite{DeGennes1963,Werthamer1963}.

In theoretical models of superconducting proximity, the role of the nature of the interface between the two materials is often neglected. When extending our considerations to an interface with limited transmission of electrons and Cooper pairs, a discontinuous jump in the order parameter arises at the interface \cite{Blonder1982-tm}. A fully transparent interface would result in a smooth variation of the order parameter across the interface, while a highly reflective interface would lead to an abrupt jump of the order parameter at the interface.

In this work we consider a crystalline superconductor-normal (SN) interface, and show that the proximity effect can be varied drastically, depending on the stacking and magnetic order of the normal metal. Our model system is Co on Ru(0001) \cite{Gabaly2007,Herve2017,Mougel2020}, in which Co can grow either in the ferromagnetic hcp phase, in registry with the crystal structure of Ru, or in a non-magnetic \textepsilon-like phase with a broad reconstruction which breaks the translational invariance of the interface. 

Co/Ru(0001) samples were prepared under ultra-high vacuum (UHV) at a base pressure of $4\times 10^{-11}$ mbar. The Ru(0001) single crystal was cleaned by cycles of annealing in oxygen and argon-ion sputtering followed by thermal annealing \cite{Herve2017}. On the atomically flat and clean surfaces, Co films were deposited from an e-beam evaporator followed by a transfer to a scanning tunneling microscope (STM) under UHV. STM measurements were performed at 30 mK with a home-built microscope \cite{Balashov2018}.

\begin{figure}[h!]
    \centering
    \includegraphics[width = 0.9\columnwidth]{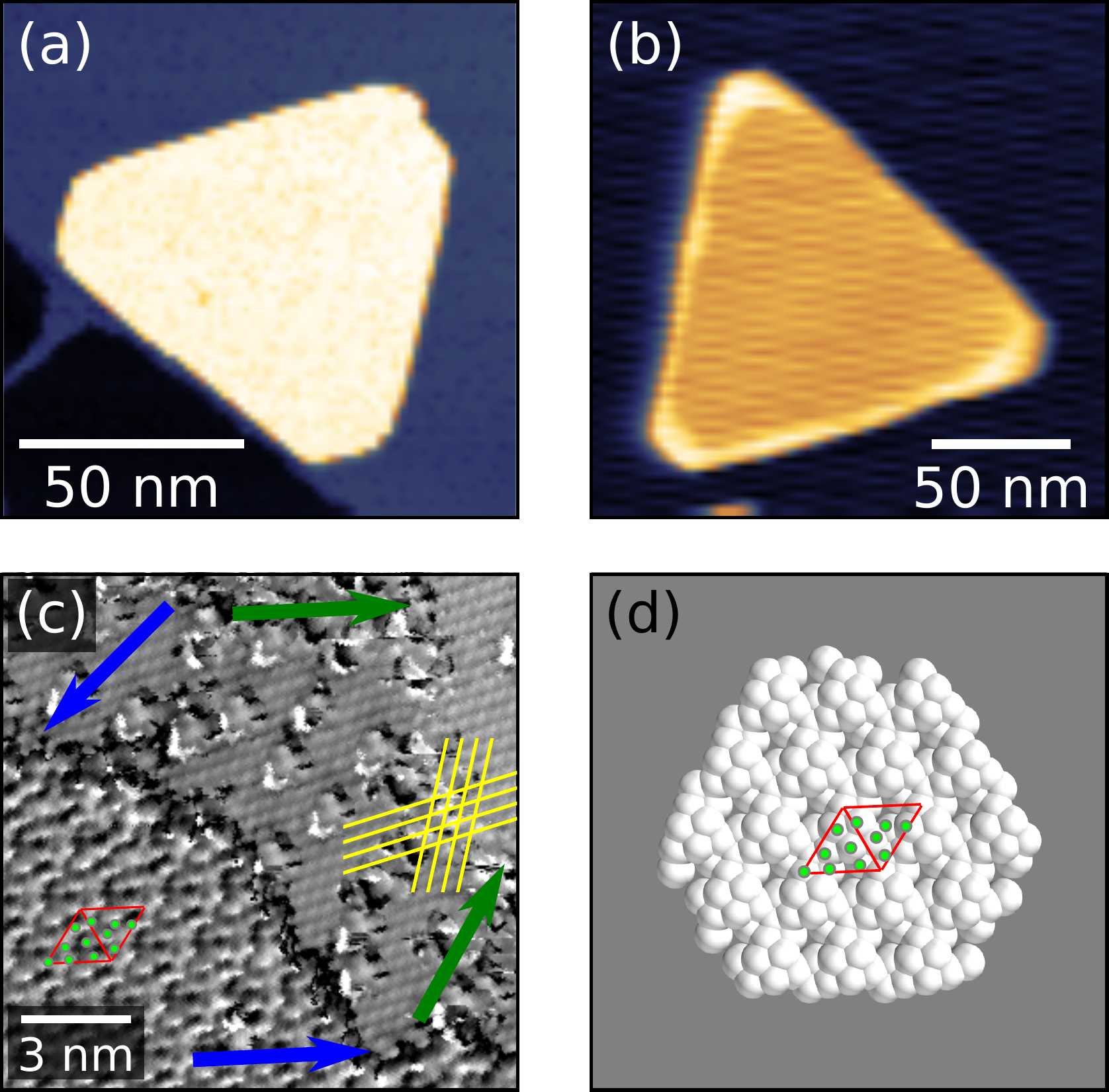}
\caption{Topographic images of the two different stackings of Co sub ML coverages on the Ru(0001) surface: (a) hcp and (b) \textepsilon ($U=$1 V, $I =$ 1 nA. (c) $dI/dz$ image of the different stackings of Co near a Ru(0001) step edge ($U=$100 mV, $I =$ 1 nA ,$z_{mod}=20~\text{pm}$. 
(d) (111) cut of \textepsilon-Co phase. The red diamond shows the unit cell and yellow lines highlight rows of Co then Ru visible in the hcp surface. Green dots show the positions of the surface Co atoms. The positions are repeated in (c) for comparison.}
\label{fig:Co_atom}
\end{figure}

In our previous work, we reported on the magnetic ground state of hcp Co on Ru(0001), which is a Bloch-type spin spiral \cite{Herve2017}. Here we show that, depending on the deposition parameters, two differently stacked phases appear. Besides the hcp stacking of the Co layer, which forms triangular islands [see Figure \ref{fig:Co_atom}(a)], islands with opposite step edge orientation (reversed triangles) can be found [see Figure \ref{fig:Co_atom}(b)]. These islands appear about 50 pm lower in the STM images. No spin contrast was observed on them down to the atomic level and no change of the spin signal could be induced by applying a magnetic field, implying a non-magnetic state for Co in these islands. Figure \ref{fig:Co_atom}(c) shows a zoomed area with atomic resolution containing both phases. It was recorded near a Ru upward step edge (green arrow) with a narrow stripe of hcp stacked Co attached to it. The crystal lattice going from Co to Ru shows no shifts, confirming the stacking (yellow lines). Next to the hcp Co, the second phase is visible separated by a phase boundary (blue arrows). The phase shows a large unit cell in the form of a $(\sqrt{10}\times\sqrt{10})$-reconstruction. The basis vectors are indicated by red arrows and have a length of 939 pm. This unit cell agrees well with the 2d unit cell of (111) bulk \textepsilon-Co of 860 pm \cite{delapena2010} as shown in Figure \ref{fig:Co_atom}(d). The positions of the atoms in that unit cell (green dots) agree qualitatively with the STM image. Note that the binary phase diagram of Co and Ru contains a phase designated as \textepsilon \cite{Koster1952}, but it differs from the phase we observe, as it describes the hcp-phase. This phase was also found for pure Co in small clusters \cite{delapena2010}. The Co atoms in the unit cell are less densely packed than in the hcp lattice. We expect that some of the atoms reside in hcp hollow sites of Ru(0001), others in the similar fcc hollow sites, as well as intermediate positions.
Further, the \textepsilon-islands display a brighter, i.e. higher, border, as can be seen in Figure \ref{fig:Co_atom}b. Note that the \textepsilon-phase is structurally rather complex, such that a simplified fcc structure will be used later in the theoretical description.

\begin{figure}[h!]
    \centering 
    \includegraphics[width = 0.9\columnwidth]{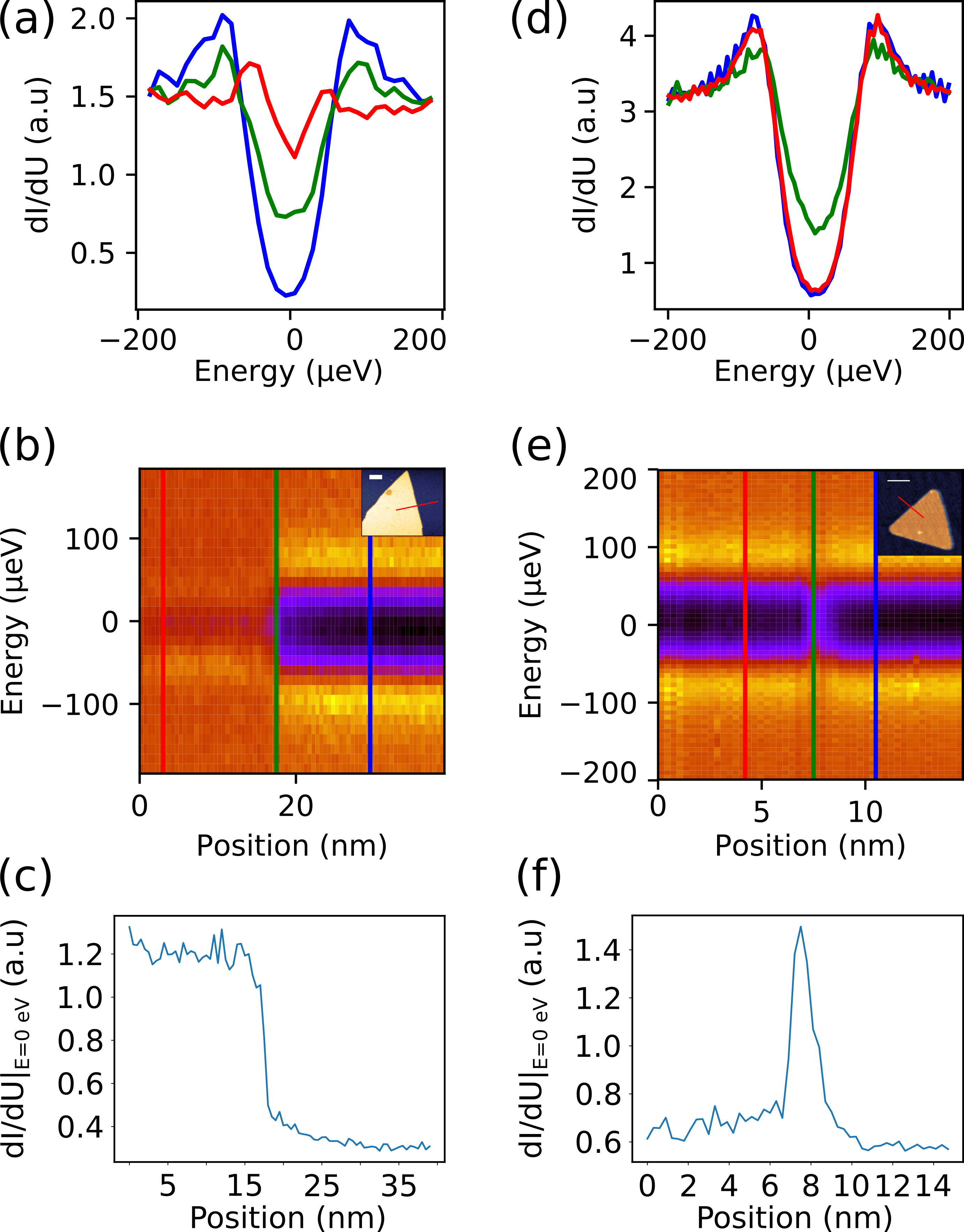}
\caption{$dI/dU$ scans of the Co overlayers on Ru. Left panels: hcp-Co. Right panels: \textepsilon-Co. (a) and (d): Individual $dI/dU$ spectra recorded on free Ru (blue), edge of the island (green) and on the bulk of the island (red). The line profiles are extracted from the $dI/dU$ spectra visible on (b) and (e). (b) and (e) : Color coded $dI/dU$ spectra plotted against lateral position of the tip crossing the island edge. The left halves of the spectra are recorded on the island, while the right halves are recorded on free Ru. Insets show the topographic images of the islands on which the spectra were recorded. The red line shows the scan trajectory of the tip. (c) and (f): $dI/dU$ value at the Fermi energy plotted against the position of the tip.}
\label{fig:dIdV}
\end{figure}

Ru has a superconducting critical temperature of $T_c=470~\text{mK}$ \cite{Finnemore1962}, i.e. is superconducting at the measurement temperature of 30 mK. To investigate the proximity effect between Ru and the Co islands, we recorded local tunneling spectra. Figure \ref{fig:dIdV}(a) and (d) show individual $dI/dU$ spectra recorded in three positions as indicated by the colour code. The insets of Figure \ref{fig:dIdV}(b) and (e) show the individual hcp and \textepsilon-islands as well as the line sections on which the spectra were taken.

Next to both islands and on the Ru substrate, the spectra (blue lines) show a superconducting gap of $\Delta=60.7\pm0.7 \mu$eV which is slightly lower than our previous measurements on bare Ru(0001) \cite{Balashov2018}. Additionally, the gap is incomplete, i.e. the differential conductance does not vanish at zero bias. 
This can be easily explained by the estimated coherence length for Ru is $\xi=\frac{\hbar v_f}{\pi\Delta}=3.4~\text{\textmu m}$. This is much larger than the average Co island size and their separation. Thus, the effect of the islands on superconductivity of the substrate is spatially averaged and the gap on the free Ru surface is consistently reduced on the whole surface, due to the proximity effect. Placing the tip on the island edge, the behavior of hcp and \textepsilon-islands is similar (green lines). The zero-bias conductance is further increased, i.e. the Cooper pair density decreased. Measuring inside the islands (red lines), however, the spectra differ dramatically. While on the \textepsilon-island the spectrum is almost identical to that of the bare Ru, on the hcp island we find a strong reduction of the gap intensity. Further, states inside the gap evolve that are slightly asymmetric 
which is not uncommon, when a magnetic metal or impurity is brought into contact with a superconductor. Figure \ref{fig:dIdV}(b) and (e) show colour coded $dI/dU$ spectra as a function of lateral displacement when going from the island (left) over the edge to the bare substrate (right). First, we note that the reduced gap on the edge of the \textepsilon-island coincides with the bright rim observed in the STM topography. Atomically resolved images of the edge (not shown) indicate that the rim consists of hcp Co. This also explains the similar spectra for the two island edges.   

Figure \ref{fig:dIdV}(b) and (e) give a more detailed view on the tunneling spectra as function of position. In both contour plots, the spectra evolve continuously when going from the substrate over the edge and into the island. Notably, the position of the coherence peaks shifts to slightly lower energies when coming close to the islands. At the same time, the $dI/dU$ signal at zero bias increases gradually over a distance of a few nm. To analyze this, we plot $dI/dU$ at zero bias as a function of position [see Figure \ref{fig:dIdV}(c) and (f)].

For the magnetic hcp island, the quasi particle density starts a gradual increase about $\approx$15 nm before the island edge, then abruptly jumps at the edge and is essentially constant on the island. The first effect can be simply explained by the dimensionality $n$ of the problem. In general, the proximity effect leads to variations of the cooper pair density in a superconductor with the function $|\psi|^2 \approx r^{-(n-1)} e^{- r/\xi}$, i.e. for a 1d problem the usual exponential decay is found, while for higher dimensions, the scattering geometry has to be considered. For a 3d situation the $1/r^2$ factor simply represents particle conservation. We do not attempt to fit the dependence: first of all $\xi$ is so large that no meaningful number can be fit on sections of a length of only few nm; and second, the dimensionality of the problem near an island should display a crossover from 2d to 3d. Essentially the same behaviour is found for the \textepsilon-island. The sudden jump, however, indicates partial transmission of electrons at the interface. In more detail, the jump on the magnetic hcp island separates the superconductor with only a weak suppression of superconductivity from the magnetic island, where almost no Cooper pairs exist. This scenario can only be observed for a rather opaque interface, decoupling the two regions. A fully transparent interface would lead to a strong reduction of the superconducting order parameter in Ru close to the island.  
In contrast, the \textepsilon-island shows a only slightly higher differential conductance at zero bias than the bare substrate. This indicates a strong proximity effect and nearly the same order parameter as the substrate. The interface must be highly transparent, and thanks to the absence of magnetic order on the island Cooper pair breaking by the exchange interaction is absent.

To gain insight into the role of interface transparency and magnetic order on the proximity effect, we have carried out density function theory (DFT) calculations combined with tight-binding (TB) simulations for the hcp stacking and the simplified fcc stacking. The DFT calculations were carried out with the FLEUR ab initio package \cite{FLEUR} which uses a FLAPW basis set to describe the electron wavefunctions \cite{Weinert2009-iw}. The exchange and correlation was computed based on the PBE approximation \cite{Perdew1992-hd}.
The muffin tin (MT) radii of Co and Ru are fixed to 2.27 and 2.40 Bohr, respectively. The plane wave basis cutoff is $K_{\mathrm{max}}=4.2$ Bohr$^{-1}$.
The atomic position of the Co and the top Ru layer were relaxed until the forces reached 0.001 Hartree/Bohr. The hcp structure is the ground state, and the fcc Co stacking is 114.54 meV/Co higher in energy (we recall that fcc Co is used as an approximant to the \textepsilon{} phase). The calculations were also extended to non-magnetic hcp and fcc stackings, which are higher in energy by 251.24 meV/Co 235.78 meV/Co, respectively.
We have computed the charge density of a symmetric monolayer of Co on Ru(0001) slabs, for which we have changed the Ru thickness from 10 to 19 layers resulting on 10 independent Ru layer for the thicker slab. 10 independent Ru layers are necessary to obtain a good approximation of the bulk Ru band structure away from the Co monolayer.

Superconducting ultra-thin films are rarely explored theoretically due to the complexity of their band structure at the Fermi level. We characterize the induced superconductivity from the Ru substrate in the Co layer using a tight-binding framework \cite{Menard2017-lg,Garnier2019-ak}. In this basis we start with a superconducting bath, described by the Bogoliubov de Gennes (BdG) equation:
\begin{eqnarray}
H^{\mathrm{Ru}} = \sum_{i} \mu^{\mathrm{Ru}}_{i} c^{\dag}_i c_i + \sum_{ij} t^{\mathrm{Ru}}_{ij} c^{\dag}_i c_j + \sum_{i} \Delta^{\mathrm{Ru}}_{ii} c^{\dag}_i c^{\dag}_{i} + h.c. \;,
\label{eq:supra}
\end{eqnarray}

\noindent where $\mu^{\mathrm{Ru}}$ is the chemical potential, $t^{\mathrm{Ru}}_{ij}$ is the hopping parameter between different sites and $\Delta^{\mathrm{Ru}}_{ii}$ is the superconducting gap energy. The BdG equation is then coupled to a normal metal overlayer which can be parametrized by a simple tight binding Hamiltonian containing only the hopping to nearest neighbors and the chemical potential:

\begin{eqnarray}
H^{\mathrm{Co}} = \sum_{i} \mu^{\mathrm{Co}}_{i} c^{\dag}_i c_i + \sum_{ij} t^{\mathrm{Co}}_{ij} c^{\dag}_i c_j + J_{\mathrm{sd}} \sum_{i} \sigma_i \mathbf{M}_i \;,
\label{eq:mag}
\end{eqnarray}

\noindent where $J_{\mathrm{sd}}$ represents the coupling between the conducting s electrons of both Ru and Co atoms and the d electrons of the Co represented by their magnetization $\mathbf{M}_i$. The total Hamiltonian $H$ is then given by $H=H^{\mathrm{Ru}}+H^{\mathrm{Co}}$ and contains one part which induces superconductivity and one part which suppresses it. In this framework, the $J_{\mathrm{sd}}$ coupling can also be interpreted as a local magnetic field which acts on each atomic site. If this local magnetic field is equal to or greater than the superconducting gap, superconductivity will not be induced in the Co layer. The $J_{\mathrm{sd}}$ can be approximated from DFT calculations by looking at the spin polarization of the exchange and correlation potential $J_{\mathrm{sd}}=\left( V^{\mathrm{xc}}_{\mathrm{up}} - V^{\mathrm{xc}}_{\mathrm{down}} \right)$. In the FLAPW basis set, this difference can be easily obtained by looking at the potential difference in the interstitial region where the basis set is composed of plane waves. This potential difference can also be understood as the polarization of the s electrons of the Co and the Ru by the d electrons of the Co.

We have checked that $J_{\mathrm{sd}}=0$ eV in case of bulk Ru which is non magnetic. In case of bulk Fe, we obtain $J_{\mathrm{sd}}=0.7$ eV which is very similar to textbook values \cite{Janak1977-kr}. In the case of Co/Ru(0001), we obtain $J_{\mathrm{sd}}=1.18\pm 0.006$ eV/Co for both stackings without spin orbit coupling (SOC). This increases to $J_{\mathrm{sd}}=1.2 \pm 0.06$ eV/Co when SOC is taken into account. This value can then be compared to the superconducting gap extracted from the measurements presented in Fig. \ref{fig:dIdV} (a) and (d), which is lower than 400 $\mu$eV. This value is two orders of magnitude smaller than the $J_{\mathrm{sd}}$ extracted from DFT, which indicates that none of the magnetic Co stacking should be superconducting on Ru. Since the \textepsilon{} phase shows a complex reconstruction, a $J_{\mathrm{sd}}$ parameter was also extracted using a $2 \times 2$ supercell containing a vacancy site. In that case $J_{\mathrm{sd}}=0.51$ eV/site. This value of $J_{\mathrm{sd}}$ for a magnetic ordered fcc phase is smaller but still remains orders of magnitude larger than $\Delta^{\mathrm{Ru}}$.

An explanation for the occurrence of superconductivity in the Co fcc-like phase on Ru calls for a quenching of the Co magnetic moments i.e. the suppression of magnetism in Co due to the surface reconstruction. To explore this possibility we have also obtained the charge density of one monolayer of non-magnetic Co(hcp) and Co(fcc) phase. Both configurations are rather close in energy and could therefore occur. Although such quenching of magnetic moment is rather rare at interfaces between 3\emph{d} transition metals on 4\emph{d} or 5\emph{d} substrates, magnetically dead layer have been described in the literature, e.g. in 2Mn/W(001) \cite{Meyer2020-pg}. The spin-dependent chemical potential and the hopping parameters in Eq.\ref{eq:supra} were parametrized via Wannier functions \cite{Marzari2012-du,Pizzi2020-hh}. We have used 6 orbitals for each Ru atom and 9 orbitals for Co in one monolayer of Co on 10 Ru layers, resulting in 69 orbitals per spin channel. The gap $\Delta$ was set to 20 meV and the Hamiltonian was diagonalized using a $8000 \times 8000$ 2-dimensional k-point grid. The value of the gap $\Delta$ is chosen much larger than the measured gap of 400 $\mu$eV, in order to limit the computing time needed to diagonalize Eq.\ref{eq:supra}, but still much smaller thant $J_{\mathrm{sd}}$, and qualitative conclusions will be drawn.

\begin{figure}[h!]
 \centering
    \includegraphics[width = \columnwidth]{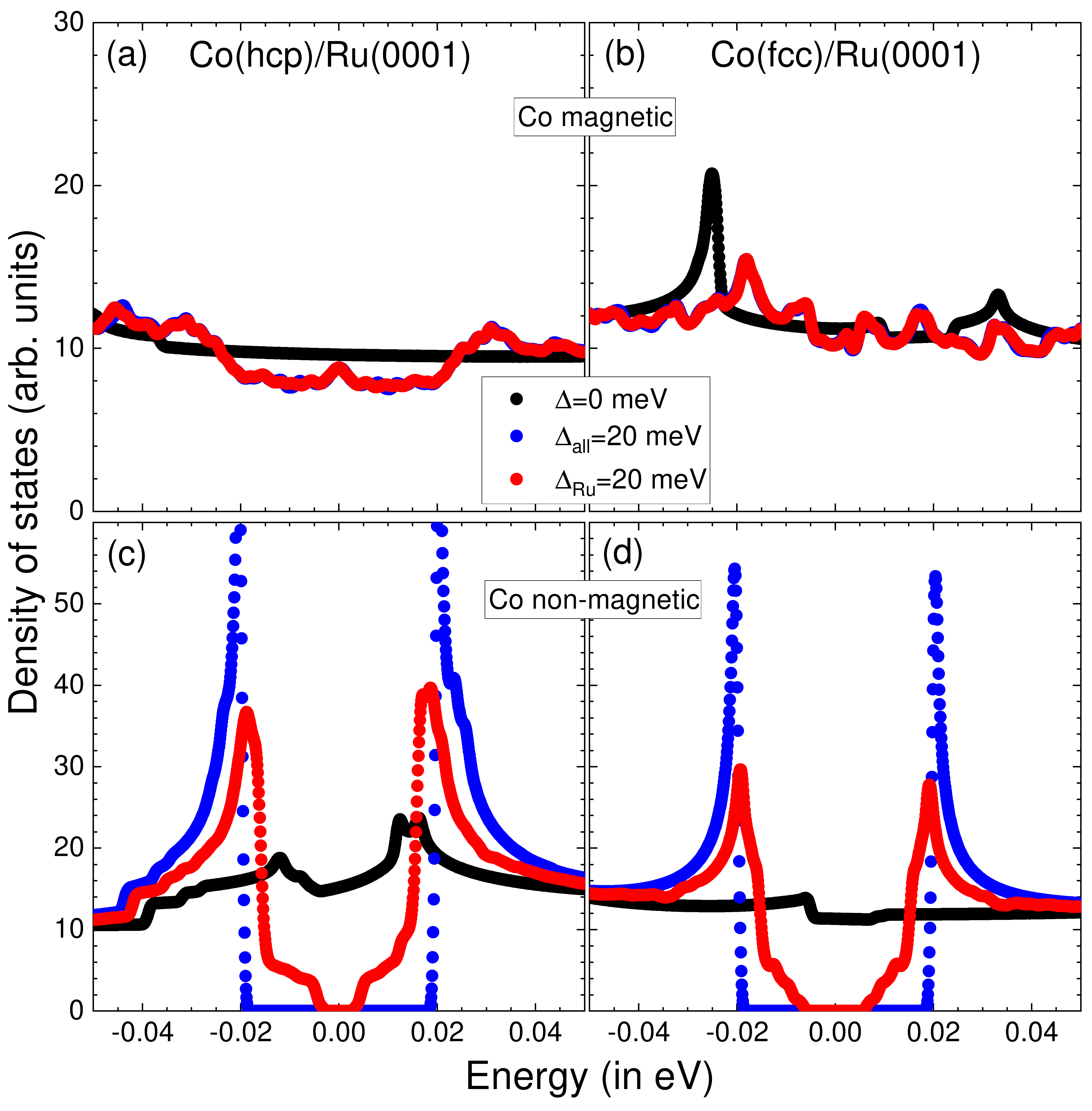}
\caption{Density of states (DOS) for non-magnetic hcp (left) and fcc (right) monolayers of Co on 10 Ru substrates atoms. 3 case are compared: the absence of superconducting gap $\Delta=0$ meV (black dots); a superconducting gap $\Delta_{\mathrm{all}}=20$ meV applied on all orbitals in both the Co and the Ru layers (blue dots); and $\Delta_{\mathrm{Ru}}=20$ meV applied only on Ru orbitals. A Gaussian smearing of 0.5 meV was applied to all DOS.}
\label{fig:DOS_SC}
\end{figure}

Figure \ref{fig:DOS_SC} summarizes the findings of our calculation. 
Firstly, we have tried to induce a superconducting gap in the Co layers when Co was magnetic as shown in Fig.\ref{fig:DOS_SC} panels (a) and (b).  The local superconducting gap was applied on the Ru orbitals only (red curve), but also extended to the Co orbitals (blue curve) as a stress test to maximize superconductivity. In both cases, the superconducting gap in Co is strongly reduced: in the case of the Co(hcp)/Ru(0001) Fig.\ref{fig:DOS_SC}(a), a small remnant gap remains (compare with the black curve in absence of superconductivity) similarly to the experiments on hcp islands. This also implies a reduced transparency of the interface.
In contrast, in the magnetic fcc phase no traces of superconductivity are found under any circumstances (see Fig.\ref{fig:DOS_SC}(b)). 

In contrast to the magnetic calculations, the non-magnetic ones in Fig. \ref{fig:DOS_SC}(c and d) always show a proximity effect. The fcc case shows a strong proximity effect in which the gap is almost completely developed in the non-magnetic Co (red curve), in agreement with the experiment on \textepsilon{} islands. The non-magnetic hcp structure shows a smaller gap in comparison to the fcc case, when only the Ru is made superconducting (red curve). This also implies reduced transparency at the interface of hcp in comparison to fcc, in analogy to the magnetic case, which explains the quenching of superconductivity at the (non-magnetic) hcp edges of \textepsilon{} islands in experiments.

In conclusion we have shown that Co grows in 2 different stackings on a Ru(0001) substrate. Co grows pseudomorphically with an hcp stacking, in which the superconductivity is strongly suppressed in Co by the combination of the magnetic order and an interface with low transparency. In the \textepsilon{} like phase, the Co monolayer exhibits a complex reconstruction which is responsible for a high transparency of the interface for Cooper pairs from the Ru substrate. Additionally, the experimentally observed absence of magnetic order suppresses Cooper pair break up.
Our study also shows that the proximity effect between a magnetic metal and a superconductor does not necessary kill superconductivity in the latter, if the interface transparency is low. Instead, the Cooper pair density shows an abrupt jump at the interface. Our results illustrate that both the interface transparency and the $J_{\mathrm{sd}}$ coupling should be studied on equal footing to explain interfacial superconductivity.

\section*{Acknowledgement}

W. Wulfhekel and L. Mougel acknowledge funding by the Deutsche Forschungsgemeinschaft (DFG) under Grant Nos. WU 349/15-1 and WU 349/16-1. H.S. Yang acknowledges funding by the Alexander-von-Humboldt Foundation. B. Dup\'e and P. Buhl acknowledge funding by the DFG under Grant No. DU 1489/3-1. B. Dup\'e and P. Buhl would like to thank S. Bl\"ugel and P. R\"ußmann for fruitful discussion. B. Dup\'e and P. Buhl acknowledge computing time from ARCHER UK National Supercomputing Service through DECI call XX, and from the Consortium d'Equipements de Calcul Intensif (FRS-FNRS Belgium GA 2.5020.11) and the Zenobe Tier-1 (Walloon Region GA 1117545).

\bibliography{citations}

\begin{thebibliography}{22}%
\makeatletter
\providecommand \@ifxundefined [1]{%
 \@ifx{#1\undefined}
}%
\providecommand \@ifnum [1]{%
 \ifnum #1\expandafter \@firstoftwo
 \else \expandafter \@secondoftwo
 \fi
}%
\providecommand \@ifx [1]{%
 \ifx #1\expandafter \@firstoftwo
 \else \expandafter \@secondoftwo
 \fi
}%
\providecommand \natexlab [1]{#1}%
\providecommand \enquote  [1]{``#1''}%
\providecommand \bibnamefont  [1]{#1}%
\providecommand \bibfnamefont [1]{#1}%
\providecommand \citenamefont [1]{#1}%
\providecommand \href@noop [0]{\@secondoftwo}%
\providecommand \href [0]{\begingroup \@sanitize@url \@href}%
\providecommand \@href[1]{\@@startlink{#1}\@@href}%
\providecommand \@@href[1]{\endgroup#1\@@endlink}%
\providecommand \@sanitize@url [0]{\catcode `\\12\catcode `\$12\catcode
  `\&12\catcode `\#12\catcode `\^12\catcode `\_12\catcode `\%12\relax}%
\providecommand \@@startlink[1]{}%
\providecommand \@@endlink[0]{}%
\providecommand \url  [0]{\begingroup\@sanitize@url \@url }%
\providecommand \@url [1]{\endgroup\@href {#1}{\urlprefix }}%
\providecommand \urlprefix  [0]{URL }%
\providecommand \Eprint [0]{\href }%
\providecommand \doibase [0]{https://doi.org/}%
\providecommand \selectlanguage [0]{\@gobble}%
\providecommand \bibinfo  [0]{\@secondoftwo}%
\providecommand \bibfield  [0]{\@secondoftwo}%
\providecommand \translation [1]{[#1]}%
\providecommand \BibitemOpen [0]{}%
\providecommand \bibitemStop [0]{}%
\providecommand \bibitemNoStop [0]{.\EOS\space}%
\providecommand \EOS [0]{\spacefactor3000\relax}%
\providecommand \BibitemShut  [1]{\csname bibitem#1\endcsname}%
\let\auto@bib@innerbib\@empty
\bibitem [{\citenamefont {Holm}\ and\ \citenamefont
  {Meissner}(1932)}]{Holm1932}%
  \BibitemOpen
  \bibfield  {author} {\bibinfo {author} {\bibfnamefont {R.}~\bibnamefont
  {Holm}}\ and\ \bibinfo {author} {\bibfnamefont {W.}~\bibnamefont
  {Meissner}},\ }\bibfield  {title} {\bibinfo {title} {Messungen mit hilfe von
  fl{\"u}ssigem helium. xiii},\ }\href {https://doi.org/10.1007/BF01340420}
  {\bibfield  {journal} {\bibinfo  {journal} {Zeitschrift f{\"u}r Physik}\
  }\textbf {\bibinfo {volume} {74}},\ \bibinfo {pages} {715} (\bibinfo {year}
  {1932})}\BibitemShut {NoStop}%
\bibitem [{\citenamefont {Cooper}(1956)}]{Cooper1956}%
  \BibitemOpen
  \bibfield  {author} {\bibinfo {author} {\bibfnamefont {L.~N.}\ \bibnamefont
  {Cooper}},\ }\bibfield  {title} {\bibinfo {title} {Bound electron pairs in a
  degenerate fermi gas},\ }\href {https://doi.org/10.1103/PhysRev.104.1189}
  {\bibfield  {journal} {\bibinfo  {journal} {Phys. Rev.}\ }\textbf {\bibinfo
  {volume} {104}},\ \bibinfo {pages} {1189} (\bibinfo {year}
  {1956})}\BibitemShut {NoStop}%
\bibitem [{\citenamefont {Bardeen}\ \emph {et~al.}(1957)\citenamefont
  {Bardeen}, \citenamefont {Cooper},\ and\ \citenamefont
  {Schrieffer}}]{Bardeen1957}%
  \BibitemOpen
  \bibfield  {author} {\bibinfo {author} {\bibfnamefont {J.}~\bibnamefont
  {Bardeen}}, \bibinfo {author} {\bibfnamefont {L.~N.}\ \bibnamefont
  {Cooper}},\ and\ \bibinfo {author} {\bibfnamefont {J.~R.}\ \bibnamefont
  {Schrieffer}},\ }\bibfield  {title} {\bibinfo {title} {Microscopic theory of
  superconductivity},\ }\href {https://doi.org/10.1103/PhysRev.106.162}
  {\bibfield  {journal} {\bibinfo  {journal} {Phys. Rev.}\ }\textbf {\bibinfo
  {volume} {106}},\ \bibinfo {pages} {162} (\bibinfo {year}
  {1957})}\BibitemShut {NoStop}%
\bibitem [{\citenamefont {{De Gennes}}\ and\ \citenamefont
  {Guyon}(1963)}]{DeGennes1963}%
  \BibitemOpen
  \bibfield  {author} {\bibinfo {author} {\bibfnamefont {P.}~\bibnamefont {{De
  Gennes}}}\ and\ \bibinfo {author} {\bibfnamefont {E.}~\bibnamefont {Guyon}},\
  }\bibfield  {title} {\bibinfo {title} {Superconductivity in “normal”
  metals},\ }\href {https://doi.org/10.1016/0031-9163(63)90401-3} {\bibfield
  {journal} {\bibinfo  {journal} {Phys. Lett.}\ }\textbf {\bibinfo {volume}
  {3}},\ \bibinfo {pages} {168} (\bibinfo {year} {1963})}\BibitemShut {NoStop}%
\bibitem [{\citenamefont {Werthamer}(1963)}]{Werthamer1963}%
  \BibitemOpen
  \bibfield  {author} {\bibinfo {author} {\bibfnamefont {N.~R.}\ \bibnamefont
  {Werthamer}},\ }\bibfield  {title} {\bibinfo {title} {Theory of the
  superconducting transition temperature and energy gap function of superposed
  metal films},\ }\href {https://doi.org/10.1103/PhysRev.132.2440} {\bibfield
  {journal} {\bibinfo  {journal} {Phys. Rev.}\ }\textbf {\bibinfo {volume}
  {132}},\ \bibinfo {pages} {2440} (\bibinfo {year} {1963})}\BibitemShut
  {NoStop}%
\bibitem [{\citenamefont {Blonder}\ \emph {et~al.}(1982)\citenamefont
  {Blonder}, \citenamefont {Tinkham},\ and\ \citenamefont
  {Klapwijk}}]{Blonder1982-tm}%
  \BibitemOpen
  \bibfield  {author} {\bibinfo {author} {\bibfnamefont {G.~E.}\ \bibnamefont
  {Blonder}}, \bibinfo {author} {\bibfnamefont {M.}~\bibnamefont {Tinkham}},\
  and\ \bibinfo {author} {\bibfnamefont {T.~M.}\ \bibnamefont {Klapwijk}},\
  }\bibfield  {title} {\bibinfo {title} {Transition from metallic to tunneling
  regimes in superconducting microconstrictions: Excess current, charge
  imbalance, and supercurrent conversion},\ }\href
  {https://doi.org/10.1103/PhysRevB.25.4515} {\bibfield  {journal} {\bibinfo
  {journal} {Phys. Rev. B}\ }\textbf {\bibinfo {volume} {25}},\ \bibinfo
  {pages} {4515} (\bibinfo {year} {1982})}\BibitemShut {NoStop}%
\bibitem [{\citenamefont {Gabaly}\ \emph {et~al.}(2007)\citenamefont {Gabaly},
  \citenamefont {Puerta}, \citenamefont {Klein}, \citenamefont {Saa},
  \citenamefont {Schmid}, \citenamefont {McCarty}, \citenamefont {Cerda},\ and\
  \citenamefont {de~la Figuera}}]{Gabaly2007}%
  \BibitemOpen
  \bibfield  {author} {\bibinfo {author} {\bibfnamefont {F.~E.}\ \bibnamefont
  {Gabaly}}, \bibinfo {author} {\bibfnamefont {J.~M.}\ \bibnamefont {Puerta}},
  \bibinfo {author} {\bibfnamefont {C.}~\bibnamefont {Klein}}, \bibinfo
  {author} {\bibfnamefont {A.}~\bibnamefont {Saa}}, \bibinfo {author}
  {\bibfnamefont {A.~K.}\ \bibnamefont {Schmid}}, \bibinfo {author}
  {\bibfnamefont {K.~F.}\ \bibnamefont {McCarty}}, \bibinfo {author}
  {\bibfnamefont {J.~I.}\ \bibnamefont {Cerda}},\ and\ \bibinfo {author}
  {\bibfnamefont {J.}~\bibnamefont {de~la Figuera}},\ }\bibfield  {title}
  {\bibinfo {title} {Structure and morphology of ultrathin co/ru(0001) films},\
  }\href@noop {} {\bibfield  {journal} {\bibinfo  {journal} {New J. Phys.}\
  }\textbf {\bibinfo {volume} {9}},\ \bibinfo {pages} {80} (\bibinfo {year}
  {2007})}\BibitemShut {NoStop}%
\bibitem [{\citenamefont {Herv{\'{e}}}\ \emph {et~al.}(2018)\citenamefont
  {Herv{\'{e}}}, \citenamefont {Dup{\'{e}}}, \citenamefont {Lopes},
  \citenamefont {B{\"{o}}ttcher}, \citenamefont {Martins}, \citenamefont
  {Balashov}, \citenamefont {Gerhard}, \citenamefont {Sinova},\ and\
  \citenamefont {Wulfhekel}}]{Herve2017}%
  \BibitemOpen
  \bibfield  {author} {\bibinfo {author} {\bibfnamefont {M.}~\bibnamefont
  {Herv{\'{e}}}}, \bibinfo {author} {\bibfnamefont {B.}~\bibnamefont
  {Dup{\'{e}}}}, \bibinfo {author} {\bibfnamefont {R.}~\bibnamefont {Lopes}},
  \bibinfo {author} {\bibfnamefont {M.}~\bibnamefont {B{\"{o}}ttcher}},
  \bibinfo {author} {\bibfnamefont {M.~D.}\ \bibnamefont {Martins}}, \bibinfo
  {author} {\bibfnamefont {T.}~\bibnamefont {Balashov}}, \bibinfo {author}
  {\bibfnamefont {L.}~\bibnamefont {Gerhard}}, \bibinfo {author} {\bibfnamefont
  {J.}~\bibnamefont {Sinova}},\ and\ \bibinfo {author} {\bibfnamefont
  {W.}~\bibnamefont {Wulfhekel}},\ }\bibfield  {title} {\bibinfo {title}
  {Stabilizing spin spirals and isolated skyrmions at low magnetic field
  exploiting vanishing magnetic anisotropy},\ }\href
  {https://doi.org/10.1038/s41467-018-03240-w} {\bibfield  {journal} {\bibinfo
  {journal} {Nat. Commun.}\ }\textbf {\bibinfo {volume} {9}},\ \bibinfo {pages}
  {1015} (\bibinfo {year} {2018})}\BibitemShut {NoStop}%
\bibitem [{\citenamefont {Mougel}\ \emph {et~al.}(2020)\citenamefont {Mougel},
  \citenamefont {Buhl}, \citenamefont {Nemoto}, \citenamefont {Balashov},
  \citenamefont {Hervé}, \citenamefont {Skolaut}, \citenamefont {Yamada},
  \citenamefont {Dupé},\ and\ \citenamefont {Wulfhekel}}]{Mougel2020}%
  \BibitemOpen
  \bibfield  {author} {\bibinfo {author} {\bibfnamefont {L.}~\bibnamefont
  {Mougel}}, \bibinfo {author} {\bibfnamefont {P.~M.}\ \bibnamefont {Buhl}},
  \bibinfo {author} {\bibfnamefont {R.}~\bibnamefont {Nemoto}}, \bibinfo
  {author} {\bibfnamefont {T.}~\bibnamefont {Balashov}}, \bibinfo {author}
  {\bibfnamefont {M.}~\bibnamefont {Hervé}}, \bibinfo {author} {\bibfnamefont
  {J.}~\bibnamefont {Skolaut}}, \bibinfo {author} {\bibfnamefont {T.~K.}\
  \bibnamefont {Yamada}}, \bibinfo {author} {\bibfnamefont {B.}~\bibnamefont
  {Dupé}},\ and\ \bibinfo {author} {\bibfnamefont {W.}~\bibnamefont
  {Wulfhekel}},\ }\bibfield  {title} {\bibinfo {title} {Instability of
  skyrmions in magnetic fields},\ }\href {https://doi.org/10.1063/5.0013488}
  {\bibfield  {journal} {\bibinfo  {journal} {Applied Physics Letters}\
  }\textbf {\bibinfo {volume} {116}},\ \bibinfo {pages} {262406} (\bibinfo
  {year} {2020})},\ \Eprint
  {https://arxiv.org/abs/https://doi.org/10.1063/5.0013488}
  {https://doi.org/10.1063/5.0013488} \BibitemShut {NoStop}%
\bibitem [{\citenamefont {Balashov}\ \emph {et~al.}(2018)\citenamefont
  {Balashov}, \citenamefont {Meyer},\ and\ \citenamefont
  {Wulfhekel}}]{Balashov2018}%
  \BibitemOpen
  \bibfield  {author} {\bibinfo {author} {\bibfnamefont {T.}~\bibnamefont
  {Balashov}}, \bibinfo {author} {\bibfnamefont {M.}~\bibnamefont {Meyer}},\
  and\ \bibinfo {author} {\bibfnamefont {W.}~\bibnamefont {Wulfhekel}},\
  }\bibfield  {title} {\bibinfo {title} {A compact ultrahigh vacuum scanning
  tunneling microscope with dilution refrigeration},\ }\href
  {https://doi.org/10.1063/1.5043636} {\bibfield  {journal} {\bibinfo
  {journal} {Review of Scientific Instruments}\ }\textbf {\bibinfo {volume}
  {89}},\ \bibinfo {pages} {113707} (\bibinfo {year} {2018})}\BibitemShut
  {NoStop}%
\bibitem [{\citenamefont {de~la Peña~O’Shea}\ \emph
  {et~al.}(2010)\citenamefont {de~la Peña~O’Shea}, \citenamefont {Moreira},
  \citenamefont {Roldán},\ and\ \citenamefont {Illas}}]{delapena2010}%
  \BibitemOpen
  \bibfield  {author} {\bibinfo {author} {\bibfnamefont {V.~A.}\ \bibnamefont
  {de~la Peña~O’Shea}}, \bibinfo {author} {\bibfnamefont {I.~d. P.~R.}\
  \bibnamefont {Moreira}}, \bibinfo {author} {\bibfnamefont {A.}~\bibnamefont
  {Roldán}},\ and\ \bibinfo {author} {\bibfnamefont {F.}~\bibnamefont
  {Illas}},\ }\bibfield  {title} {\bibinfo {title} {Electronic and magnetic
  structure of bulk cobalt: The α, β, and ε-phases from density functional
  theory calculations},\ }\href {https://doi.org/10.1063/1.3458691} {\bibfield
  {journal} {\bibinfo  {journal} {The Journal of Chemical Physics}\ }\textbf
  {\bibinfo {volume} {133}},\ \bibinfo {pages} {024701} (\bibinfo {year}
  {2010})},\ \Eprint {https://arxiv.org/abs/https://doi.org/10.1063/1.3458691}
  {https://doi.org/10.1063/1.3458691} \BibitemShut {NoStop}%
\bibitem [{\citenamefont {K\"oster}\ and\ \citenamefont
  {Horn}(1952)}]{Koster1952}%
  \BibitemOpen
  \bibfield  {author} {\bibinfo {author} {\bibfnamefont {W.}~\bibnamefont
  {K\"oster}}\ and\ \bibinfo {author} {\bibfnamefont {E.}~\bibnamefont
  {Horn}},\ }\bibfield  {title} {\bibinfo {title} {Zustandsbild und
  gitterkonstanten der legierungen des kobalts mit rhenium, ruthenium, osmium,
  rhodium und iridium},\ }\href@noop {} {\bibfield  {journal} {\bibinfo
  {journal} {Z. f. Metallk.}\ }\textbf {\bibinfo {volume} {43}},\ \bibinfo
  {pages} {444} (\bibinfo {year} {1952})}\BibitemShut {NoStop}%
\bibitem [{\citenamefont {Finnemore}\ and\ \citenamefont
  {Mapother}(1962)}]{Finnemore1962}%
  \BibitemOpen
  \bibfield  {author} {\bibinfo {author} {\bibfnamefont {D.~K.}\ \bibnamefont
  {Finnemore}}\ and\ \bibinfo {author} {\bibfnamefont {D.~E.}\ \bibnamefont
  {Mapother}},\ }\bibfield  {title} {\bibinfo {title} {Absence of an isotope
  effect in superconducting ruthenium},\ }\href
  {https://doi.org/10.1103/PhysRevLett.9.288} {\bibfield  {journal} {\bibinfo
  {journal} {Phys. Rev. Lett.}\ }\textbf {\bibinfo {volume} {9}},\ \bibinfo
  {pages} {288} (\bibinfo {year} {1962})}\BibitemShut {NoStop}%
\bibitem [{FLE()}]{FLEUR}%
  \BibitemOpen
  \href@noop {} {}\bibinfo {howpublished} {www.flapw.de}\BibitemShut {NoStop}%
\bibitem [{\citenamefont {Weinert}\ \emph {et~al.}(2009)\citenamefont
  {Weinert}, \citenamefont {Schneider}, \citenamefont {Podloucky},\ and\
  \citenamefont {Redinger}}]{Weinert2009-iw}%
  \BibitemOpen
  \bibfield  {author} {\bibinfo {author} {\bibfnamefont {M.}~\bibnamefont
  {Weinert}}, \bibinfo {author} {\bibfnamefont {G.}~\bibnamefont {Schneider}},
  \bibinfo {author} {\bibfnamefont {R.}~\bibnamefont {Podloucky}},\ and\
  \bibinfo {author} {\bibfnamefont {J.}~\bibnamefont {Redinger}},\ }\bibfield
  {title} {\bibinfo {title} {{FLAPW}: applications and implementations},\
  }\href {https://doi.org/10.1088/0953-8984/21/8/084201} {\bibfield  {journal}
  {\bibinfo  {journal} {J. Phys. C: Solid State Phys.}\ }\textbf {\bibinfo
  {volume} {21}},\ \bibinfo {pages} {84201} (\bibinfo {year}
  {2009})}\BibitemShut {NoStop}%
\bibitem [{\citenamefont {Perdew}\ \emph {et~al.}(1992)\citenamefont {Perdew},
  \citenamefont {Chevary}, \citenamefont {Vosko}, \citenamefont {Jackson},
  \citenamefont {Pederson}, \citenamefont {Singh},\ and\ \citenamefont
  {Fiolhais}}]{Perdew1992-hd}%
  \BibitemOpen
  \bibfield  {author} {\bibinfo {author} {\bibfnamefont {J.~P.}\ \bibnamefont
  {Perdew}}, \bibinfo {author} {\bibfnamefont {J.~A.}\ \bibnamefont {Chevary}},
  \bibinfo {author} {\bibfnamefont {S.~H.}\ \bibnamefont {Vosko}}, \bibinfo
  {author} {\bibfnamefont {K.~A.}\ \bibnamefont {Jackson}}, \bibinfo {author}
  {\bibfnamefont {M.~R.}\ \bibnamefont {Pederson}}, \bibinfo {author}
  {\bibfnamefont {D.~J.}\ \bibnamefont {Singh}},\ and\ \bibinfo {author}
  {\bibfnamefont {C.}~\bibnamefont {Fiolhais}},\ }\bibfield  {title} {\bibinfo
  {title} {Atoms, molecules, solids, and surfaces: Applications of the
  generalized gradient approximation for exchange and correlation},\ }\href
  {https://doi.org/10.1103/PhysRevB.46.6671} {\bibfield  {journal} {\bibinfo
  {journal} {Phys. Rev. B}\ }\textbf {\bibinfo {volume} {46}},\ \bibinfo
  {pages} {6671} (\bibinfo {year} {1992})}\BibitemShut {NoStop}%
\bibitem [{\citenamefont {M{\'e}nard}\ \emph {et~al.}(2017)\citenamefont
  {M{\'e}nard}, \citenamefont {Guissart}, \citenamefont {Brun}, \citenamefont
  {Leriche}, \citenamefont {Trif}, \citenamefont {Debontridder}, \citenamefont
  {Demaille}, \citenamefont {Roditchev}, \citenamefont {Simon},\ and\
  \citenamefont {Cren}}]{Menard2017-lg}%
  \BibitemOpen
  \bibfield  {author} {\bibinfo {author} {\bibfnamefont {G.~C.}\ \bibnamefont
  {M{\'e}nard}}, \bibinfo {author} {\bibfnamefont {S.}~\bibnamefont
  {Guissart}}, \bibinfo {author} {\bibfnamefont {C.}~\bibnamefont {Brun}},
  \bibinfo {author} {\bibfnamefont {R.~T.}\ \bibnamefont {Leriche}}, \bibinfo
  {author} {\bibfnamefont {M.}~\bibnamefont {Trif}}, \bibinfo {author}
  {\bibfnamefont {F.}~\bibnamefont {Debontridder}}, \bibinfo {author}
  {\bibfnamefont {D.}~\bibnamefont {Demaille}}, \bibinfo {author}
  {\bibfnamefont {D.}~\bibnamefont {Roditchev}}, \bibinfo {author}
  {\bibfnamefont {P.}~\bibnamefont {Simon}},\ and\ \bibinfo {author}
  {\bibfnamefont {T.}~\bibnamefont {Cren}},\ }\bibfield  {title} {\bibinfo
  {title} {Two-dimensional topological superconductivity in {Pb/Co/Si(111})},\
  }\href {https://doi.org/10.1038/s41467-017-02192-x} {\bibfield  {journal}
  {\bibinfo  {journal} {Nat. Commun.}\ }\textbf {\bibinfo {volume} {8}},\
  \bibinfo {pages} {2040} (\bibinfo {year} {2017})}\BibitemShut {NoStop}%
\bibitem [{\citenamefont {Garnier}\ \emph {et~al.}(2019)\citenamefont
  {Garnier}, \citenamefont {Mesaros},\ and\ \citenamefont
  {Simon}}]{Garnier2019-ak}%
  \BibitemOpen
  \bibfield  {author} {\bibinfo {author} {\bibfnamefont {M.}~\bibnamefont
  {Garnier}}, \bibinfo {author} {\bibfnamefont {A.}~\bibnamefont {Mesaros}},\
  and\ \bibinfo {author} {\bibfnamefont {P.}~\bibnamefont {Simon}},\ }\bibfield
   {title} {\bibinfo {title} {Topological superconductivity with deformable
  magnetic skyrmions},\ }\href {https://doi.org/10.1038/s42005-019-0226-5}
  {\bibfield  {journal} {\bibinfo  {journal} {Communications Physics}\ }\textbf
  {\bibinfo {volume} {2}},\ \bibinfo {pages} {126} (\bibinfo {year}
  {2019})}\BibitemShut {NoStop}%
\bibitem [{\citenamefont {Janak}(1977)}]{Janak1977-kr}%
  \BibitemOpen
  \bibfield  {author} {\bibinfo {author} {\bibfnamefont {J.~F.}\ \bibnamefont
  {Janak}},\ }\bibfield  {title} {\bibinfo {title} {Uniform susceptibilities of
  metallic elements},\ }\href {https://doi.org/10.1103/PhysRevB.16.255}
  {\bibfield  {journal} {\bibinfo  {journal} {Phys. Rev. B}\ }\textbf {\bibinfo
  {volume} {16}},\ \bibinfo {pages} {255} (\bibinfo {year} {1977})}\BibitemShut
  {NoStop}%
\bibitem [{\citenamefont {Meyer}\ \emph {et~al.}(2020)\citenamefont {Meyer},
  \citenamefont {Schmitt}, \citenamefont {Vogt}, \citenamefont {Bode},\ and\
  \citenamefont {Heinze}}]{Meyer2020-pg}%
  \BibitemOpen
  \bibfield  {author} {\bibinfo {author} {\bibfnamefont {S.}~\bibnamefont
  {Meyer}}, \bibinfo {author} {\bibfnamefont {M.}~\bibnamefont {Schmitt}},
  \bibinfo {author} {\bibfnamefont {M.}~\bibnamefont {Vogt}}, \bibinfo {author}
  {\bibfnamefont {M.}~\bibnamefont {Bode}},\ and\ \bibinfo {author}
  {\bibfnamefont {S.}~\bibnamefont {Heinze}},\ }\bibfield  {title} {\bibinfo
  {title} {Dead magnetic layers at the interface: Moment quenching through
  hybridization and frustration},\ }\href
  {https://doi.org/10.1103/PhysRevResearch.2.012075} {\bibfield  {journal}
  {\bibinfo  {journal} {Physical Review Research}\ }\textbf {\bibinfo {volume}
  {2}},\ \bibinfo {pages} {012075} (\bibinfo {year} {2020})}\BibitemShut
  {NoStop}%
\bibitem [{\citenamefont {Marzari}\ \emph {et~al.}(2012)\citenamefont
  {Marzari}, \citenamefont {Mostofi}, \citenamefont {Yates}, \citenamefont
  {Souza},\ and\ \citenamefont {Vanderbilt}}]{Marzari2012-du}%
  \BibitemOpen
  \bibfield  {author} {\bibinfo {author} {\bibfnamefont {N.}~\bibnamefont
  {Marzari}}, \bibinfo {author} {\bibfnamefont {A.~A.}\ \bibnamefont
  {Mostofi}}, \bibinfo {author} {\bibfnamefont {J.~R.}\ \bibnamefont {Yates}},
  \bibinfo {author} {\bibfnamefont {I.}~\bibnamefont {Souza}},\ and\ \bibinfo
  {author} {\bibfnamefont {D.}~\bibnamefont {Vanderbilt}},\ }\bibfield  {title}
  {\bibinfo {title} {Maximally localized wannier functions: Theory and
  applications},\ }\href {https://doi.org/10.1103/RevModPhys.84.1419}
  {\bibfield  {journal} {\bibinfo  {journal} {Rev. Mod. Phys.}\ }\textbf
  {\bibinfo {volume} {84}},\ \bibinfo {pages} {1419} (\bibinfo {year}
  {2012})}\BibitemShut {NoStop}%
\bibitem [{\citenamefont {Pizzi}\ \emph {et~al.}(2020)\citenamefont {Pizzi},
  \citenamefont {Vitale}, \citenamefont {Arita}, \citenamefont {Bl{\"u}gel},
  \citenamefont {Freimuth}, \citenamefont {G{\'e}ranton}, \citenamefont
  {Gibertini}, \citenamefont {Gresch}, \citenamefont {Johnson}, \citenamefont
  {Iba{\~n}ez-Azpiroz}, \citenamefont {Lee}, \citenamefont {Lihm},
  \citenamefont {Marchand}, \citenamefont {Marrazzo}, \citenamefont
  {Mokrousov}, \citenamefont {Mustafa}, \citenamefont {Nohara}, \citenamefont
  {Nomura}, \citenamefont {Paulatto}, \citenamefont {Ponc{\'e}}, \citenamefont
  {Ponweiser}, \citenamefont {Qiao}, \citenamefont {Th{\"o}le}, \citenamefont
  {Tsirkin}, \citenamefont {Wierzbowska}, \citenamefont {Marzari},
  \citenamefont {Vanderbilt}, \citenamefont {Souza}, \citenamefont {Mostofi},\
  and\ \citenamefont {Yates}}]{Pizzi2020-hh}%
  \BibitemOpen
  \bibfield  {author} {\bibinfo {author} {\bibfnamefont {G.}~\bibnamefont
  {Pizzi}}, \bibinfo {author} {\bibfnamefont {V.}~\bibnamefont {Vitale}},
  \bibinfo {author} {\bibfnamefont {R.}~\bibnamefont {Arita}}, \bibinfo
  {author} {\bibfnamefont {S.}~\bibnamefont {Bl{\"u}gel}}, \bibinfo {author}
  {\bibfnamefont {F.}~\bibnamefont {Freimuth}}, \bibinfo {author}
  {\bibfnamefont {G.}~\bibnamefont {G{\'e}ranton}}, \bibinfo {author}
  {\bibfnamefont {M.}~\bibnamefont {Gibertini}}, \bibinfo {author}
  {\bibfnamefont {D.}~\bibnamefont {Gresch}}, \bibinfo {author} {\bibfnamefont
  {T.}~\bibnamefont {Johnson}, \bibfnamefont {Charles~andKoretsune}}, \bibinfo
  {author} {\bibfnamefont {J.}~\bibnamefont {Iba{\~n}ez-Azpiroz}}, \bibinfo
  {author} {\bibfnamefont {H.}~\bibnamefont {Lee}}, \bibinfo {author}
  {\bibfnamefont {J.-M.}\ \bibnamefont {Lihm}}, \bibinfo {author}
  {\bibfnamefont {D.}~\bibnamefont {Marchand}}, \bibinfo {author}
  {\bibfnamefont {A.}~\bibnamefont {Marrazzo}}, \bibinfo {author}
  {\bibfnamefont {Y.}~\bibnamefont {Mokrousov}}, \bibinfo {author}
  {\bibfnamefont {J.~I.}\ \bibnamefont {Mustafa}}, \bibinfo {author}
  {\bibfnamefont {Y.}~\bibnamefont {Nohara}}, \bibinfo {author} {\bibfnamefont
  {Y.}~\bibnamefont {Nomura}}, \bibinfo {author} {\bibfnamefont
  {L.}~\bibnamefont {Paulatto}}, \bibinfo {author} {\bibfnamefont
  {S.}~\bibnamefont {Ponc{\'e}}}, \bibinfo {author} {\bibfnamefont
  {T.}~\bibnamefont {Ponweiser}}, \bibinfo {author} {\bibfnamefont
  {J.}~\bibnamefont {Qiao}}, \bibinfo {author} {\bibfnamefont {F.}~\bibnamefont
  {Th{\"o}le}}, \bibinfo {author} {\bibfnamefont {S.~S.}\ \bibnamefont
  {Tsirkin}}, \bibinfo {author} {\bibfnamefont {M.}~\bibnamefont
  {Wierzbowska}}, \bibinfo {author} {\bibfnamefont {N.}~\bibnamefont
  {Marzari}}, \bibinfo {author} {\bibfnamefont {D.}~\bibnamefont {Vanderbilt}},
  \bibinfo {author} {\bibfnamefont {I.}~\bibnamefont {Souza}}, \bibinfo
  {author} {\bibfnamefont {A.~A.}\ \bibnamefont {Mostofi}},\ and\ \bibinfo
  {author} {\bibfnamefont {J.~R.}\ \bibnamefont {Yates}},\ }\bibfield  {title}
  {\bibinfo {title} {Wannier90 as a community code: new features and
  applications},\ }\href {https://doi.org/10.1088/1361-648X/ab51ff} {\bibfield
  {journal} {\bibinfo  {journal} {J. Phys. Condens. Matter}\ }\textbf {\bibinfo
  {volume} {32}},\ \bibinfo {pages} {165902} (\bibinfo {year}
  {2020})}\BibitemShut {NoStop}%
\end{thebibliography}%

\end{document}